# Molecular Insights into the Mechanical Properties of Polymer–Fullerene Bulk Heterojunctions for Organic Photovoltaic Applications

*Yuta Yoshimoto,\* Sou Sugiyama, Shuntaro Shimada, Toshihiro Kaneko, Shu Takagi, and Ikuya Kinefuchi*

Department of Mechanical Engineering, The University of Tokyo, 7-3-1 Hongo, Bunkyo-ku, Tokyo 113-8656, Japan

## ABSTRACT

We investigate the mechanical properties of π-conjugated polymeric materials composed of regioregular poly(3-hexylthiophene) (P3HT) and fullerene $C_{60}$ using coarse-grained molecular dynamics simulations. Specifically, we perform tensile simulations of P3HT:$C_{60}$ composites with varied degrees of polymerization and $C_{60}$ mass fractions to obtain their stress–strain responses. Decomposition of stress tensor into kinetic energy and virial contributions indicates that the tensile moduli of the pure P3HT samples are greatly dependent on non-bonded interactions and on bonded interactions associated with bond-stretching, while the addition of $C_{60}$ leads to an increase in the tensile modulus originating from enhanced non-bonded interactions associated with $C_{60}$. Additionally, the tensile strength of the P3HT:$C_{60}$ samples correlates well with molecular chain entanglements, which are characterized by the average number of kinks per chain obtained from primitive path analysis. We also find that the upper and lower yield points characterizing strain softening become more pronounced with increasing $C_{60}$ mass fraction. Persistent homology analysis indicates that the emergence of the yield points correlates well with the coalescence of microvoids in the course of tensile deformation, resulting in the generation of larger voids. These results provide a fundamental understanding of the molecular determinants of the mechanical properties of π-conjugated polymer–fullerene composites, which can also help to interpret and predict the mechanical properties of other polymer composites containing fullerene.

## 1. INTRODUCTION

Organic solar cells (OSCs) are lightweight, flexible, and stretchable,[1–3] and they possess a diversity of molecular design that permits highly precise control over electronic, optical, and redox properties compared with their inorganic counterparts. Additionally, solution-processing protocols allow for large-scale and cost-effective manufacturing of OSCs via roll-to-roll printing processes.[4,5] Bulk heterojunction (BHJ) OSCs[6,7] are commercially viable devices that achieve deformability with acceptable power conversion efficiencies (PCEs). The active layer of a BHJ OSC consists of bicontinuous interpenetrating networks formed by electron-donating and electron-accepting materials, where the structural length scales from each domain to donor–acceptor interfaces should be on the order of 10 nm for efficient exciton dissociation at the interfaces. As such, the PCEs of BHJ OSCs depend not only on the energy-level alignments of donor/acceptor molecules but also on their chemical nature and the BHJ morphology subject to thin-film processing protocols such as annealing temperature and solvent evaporation rate.[8,9] Historically, OSCs have been shown to have the drawback of low PCEs, and many studies have focused on addressing this issue, resulting in PCEs of > 15% for single junction cells[10] and > 17% for tandem cells.[11] Meanwhile, much less work has been devoted to enhancing the mechanical properties of OSCs; however, there is now increasing interest in this area,[12–17] and the mechanical properties of traditional polymers have been well studied.[18,19] To produce flexible and stretchable OSC devices with a high yield using roll-to-roll printing processes, the mechanical properties of BHJ active layers should be investigated and enhanced alongside their electronic properties.

Experimentally, the mechanical properties of π-conjugated materials can be measured using a range of techniques, including pull testing,[20] nanoindentation,[21] and film-on-elastomer methods.[22–24] For example, the elastic modulus and yield point of a thin film can be measured using buckling-based film-on-elastomer techniques,[22,24] and strain at fracture can be approximated by the strain at which the first crack appears when the film is bonded to an elastomer,[23] while more rigorous methods have been proposed for evaluating fracture behavior by observing the propagation of cracks and microvoids using optical microscopy.[25,26] Additionally, double-cantilever-beam and four-point-bending experiments are suitable approaches for measuring the adhesion and cohesion of thin films.[15,27] Direct pull testing of a sample is also still an attractive option because it unambiguously provides the stress–strain response over a broad strain range. However, laboratory-scale synthesis can often only produce a few hundred milligrams of π-conjugated materials, and the difficulty of manipulating free-standing films with thickness of a few hundred nanometers limits the application of this technique. The stress–strain responses of π-conjugated materials provide useful information regarding their mechanical responses to external perturbation; however, molecular insight into their mechanical responses remains elusive. Therefore, to better understand the mechanical properties of these materials, it is desirable to analyze their stress–strain responses in relation to the hierarchical interplay between molecular structure, molecular conformation, packing structure, and BHJ morphology.

Molecular dynamics (MD) simulation[28,29] is a suitable method for analyzing the mechanical properties of polymeric materials from a microscopic perspective. The inherent mechanical properties of π-conjugated materials can be observed by performing *in silico* pull tests via MD simulations while retaining atomistic information.[30–32] However, all-atom MD simulations are computationally expensive because they explicitly deal with fine-grained atomistic motions according to Newton's law. Specifically, in molecular simulations for the mechanical responses of π-conjugated polymers, it is often necessary to set the degree of polymerization at a high enough level to observe the elastic behavior of the materials and the entangled behavior of the molecular chains.[30] In this regard, all-atom MD simulations are not necessarily the ideal means of investigating the mechanical properties of π-conjugated materials.

In the present study, we aim to obtain molecular insight into the mechanical properties of π-conjugated polymer–fullerene composites using coarse-grained molecular dynamics (CGMD) simulations. In CGMD simulations, multiple atoms/molecules are regarded as a single coarse-grained (CG) particle,[33–35] allowing for the simulation of much larger spatiotemporal-scale phenomena than can be achieved with all-atom MD simulations.[36–39] We select composites of regioregular poly(3-hexylthiophene) (P3HT) and fullerene $C_{60}$ as our research samples since they are widely recognized as prototypical composites for investigation of the active layers of BHJ OSCs.[40] We perform tensile simulations of P3HT:$C_{60}$ composites with varied degrees of polymerization (DPs) and $C_{60}$ mass fractions to obtain their stress–strain responses. Specifically, we focus on tensile modulus and yield behavior characterizing elastoplastic behavior, as well as tensile strength defined as maximum stress in a stress–strain curve. Decomposition of stress tensor into kinetic energy and virial contributions indicates that the tensile moduli of pure P3HT samples are highly dependent on non-bonded interactions and on bonded interactions associated with bond-stretching, while the addition of $C_{60}$ leads to an increase in the tensile modulus originating from enhanced non-bonded interactions associated with $C_{60}$. Additionally, we find that the tensile strengths of P3HT:$C_{60}$ samples correlate well with molecular chain entanglements, which are quantified via primitive path (PP) analysis.[41–44] Finally, we find that the upper and lower yield points characterizing strain softening become more pronounced as the $C_{60}$ mass fraction increases. The emergence of the yield points correlates well with the coalescence of microvoids in the course of tensile deformation, which is evaluated using persistent homology analysis.[45,46] Taken together, the present study characterizes elastoplastic behavior at a small strain regime and tensile strength at a large strain regime for P3HT:$C_{60}$ samples with varied DPs and $C_{60}$ mass fractions, providing a comprehensive understanding of the relationship between their mechanical properties and molecular-scale features over a broad strain range.

## 2. MOLECULAR DYNAMICS SYSTEMS

### 2.1. Coarse-Grained Models for P3HT:$C_{60}$ Samples.

As highlighted by Tummala et al.,[30] P3HT oligomers containing at least 50 monomer units are required to observe elastic behavior in MD simulations, while even longer chains are required for appropriate representation of molecular chain entanglements. Additionally, it is important to maintain a sufficiently large number of molecular chains to accommodate the inherent inhomogeneities of BHJs and to avoid the chains becoming entangled with their own periodic images under periodic boundary conditions.

In the present study, to deal with regioregular P3HT molecular chains with DP ≥ 50 while maintaining a sufficient number of molecular chains, we employ CGMD simulations[36,47–49] with a P3HT monomer unit and fullerene $C_{60}$ represented by three CG particles and a single CG particle, respectively, as shown in Figure 1a. Here, a thiophene ring is represented by a single CG particle ($P_1$) and a hexyl side chain is represented by two CG particles ($P_2$ and $P_3$). This CG force field[47,48] was originally developed using the iterative Boltzmann inversion method,[50,51] allowing for reproducing the structural characteristics of atomistic systems such as radial, bond, angle, and dihedral distribution functions at multiple thermodynamic-state points. Later, Root et al.[36] verified that three-site representation of a P3HT monomer unit is suitable for obtaining

P3HT equilibrium density and tensile modulus values that are in better agreement with experimental counterparts than those obtained from one-site representation of the unit.[52]

We define non-bonded interactions among $P_1$, $P_2$, and $P_3$ in a form of the Lennard-Jones (9-6) potential, and those associated with fullerene ($P_4$) in a form of the Lennard-Jones (12-6) potential with distance shifted by a certain value. The bonded interactions for P3HT include bond-stretching, angle-bending, and dihedral and improper potentials. Comprehensive descriptions of the force field are available in previous studies.[36,49] We vary DP between 50 and 150 with 300–900 molecular chains, while the $C_{60}$ mass fraction is varied between 0% and 70%, as summarized in Table 1. Periodic boundary conditions are imposed upon all the directions. All the simulations are performed with a time step of 4 fs, and all the simulations and visualizations are performed using the Large-scale Atomic/Molecular Massively Parallel Simulator (LAMMPS)[53] and Visual Molecular Dynamics (VMD).[54]

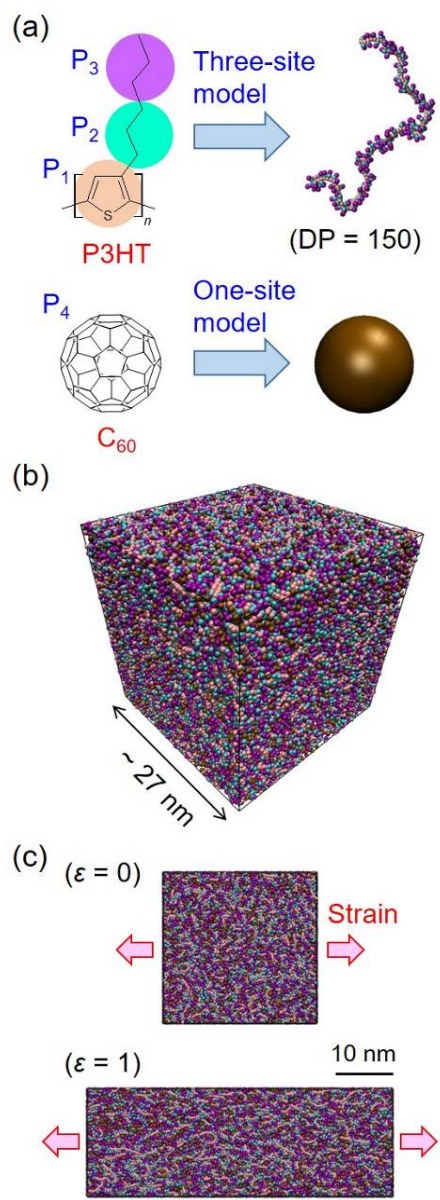

**Figure 1.** (a) A regioregular P3HT monomer unit represented by three CG particles ($P_1$, $P_2$, and $P_3$), where a thiophene ring and a hexyl side chain are represented by one ($P_1$) and two ($P_2$ and $P_3$) CG particles. Fullerene $C_{60}$ is represented by a single CG particle ($P_4$). (b) A representative system composed of 50 wt% P3HT (DP = 150) and 50 wt% $C_{60}$ at 300 K and 1 atm. (c) Uniaxial tensile simulations with an engineering strain rate ($\dot{\varepsilon}$) of $10^{-4}$ ps$^{-1}$, where the pressure perpendicular to the tensile direction is maintained at 1 atm.

**Table 1. System Descriptions for the Annealed Samples**

| DP[a] | # of chains | $C_{60}$ mass fraction (wt%) | # of $C_{60}$ | Density (g/cm$^3$)[b] | Tensile modulus (GPa)[c] |
|---|---|---|---|---|---|
| 50 | 900 | 0 | 0 | 0.9987 ± 0.0005 | 1.08 ± 0.04 |
|  |  | 30 | 4450 | 1.1468 ± 0.0004 | 1.56 ± 0.07 |
|  |  | 50 | 10383 | 1.2586 ± 0.0004 | 1.96 ± 0.02 |
|  |  | 70 | 24228 | 1.3742 ± 0.0003 | 2.62 ± 0.02 |
| 100 | 450 | 0 | 0 | 0.9982 ± 0.0005 | 1.03 ± 0.03 |
|  |  | 30 | 4450 | 1.1460 ± 0.0004 | 1.51 ± 0.02 |
|  |  | 50 | 10383 | 1.2576 ± 0.0004 | 1.97 ± 0.07 |
|  |  | 70 | 24228 | 1.3735 ± 0.0003 | 2.60 ± 0.02 |
| 150 | 300 | 0 | 0 | 0.9978 ± 0.0005 | 1.06 ± 0.05 |
|  |  | 30 | 4450 | 1.1447 ± 0.0004 | 1.52 ± 0.13 |
|  |  | 50 | 10383 | 1.2571 ± 0.0004 | 1.98 ± 0.03 |
|  |  | 70 | 24228 | 1.3731 ± 0.0003 | 2.59 ± 0.01 |

[a] DP refers to degree of polymerization.
[b] Densities are calculated at 300 K and 1 atm, where each uncertainty represents a standard deviation of the fluctuating density during isothermal–isobaric (NPT) MD simulations.
[c] Each uncertainty represents a standard deviation obtained from three independent simulations.

### 2.2. Construction of P3HT:$C_{60}$ Composite Systems.

For each case presented in Table 1, parallel stacks of extended P3HT chains with sufficient empty space between them are initially placed in a large simulation box using Moltemplate,[55] and fullerene molecules are placed in the empty space, such that an initial density of the mixture is approximately 0.01 g/cm$^3$. The sample, with non-bonded interaction parameters lowered to 10% of their condensed-phase values, is then relaxed at 600 K using a Langevin thermostat[56] with a damping factor of 180 fs. Combining this reduction of the non-bonded interaction parameters with the initial low density ($\approx$ 0.01 g/cm$^3$) mimics P3HT molecular chains and C$_{60}$ molecules in a good solvent without explicitly treating the solvent molecules, thus allowing them to explore the configuration space efficiently. This initial relaxation process lasts for 400–480 ns (depending on the conditions in Table 1), ensuring the convergences of the total potential energy, radii of gyration, and end-to-end distances of the molecular chains.

Subsequently, the non-bonded interaction parameters are gradually increased to their condensed-phase values at a temperature of 600 K and pressure of 1 atm using a Nosé–Hoover style barostat and thermostat,[57] with temperature and pressure damping parameters of 400 and 4000 fs, respectively. This gradual increase in the non-bonded parameters mimics the film densification that results from solvent evaporation in a standard coating procedure.[36] The film densification process lasts for 440–2400 ns (depending on the conditions in Table 1). Due to its slow diffusion, the addition of C$_{60}$ typically leads to a longer period of relaxation being required for the convergences of the total potential energy, radii of gyration, and end-to-end distances of the molecular chains.

Following the densification process at 600 K, the system is gradually cooled to 300 K at a rate of 5 K/ns at 1 atm, with relaxation simulations carried out at each 20-K interval. These intermediate simulations are performed to confirm the convergences of the total potential energy, radii of gyration, and end-to-end distances of the molecular chains, allowing us to obtain a sufficiently relaxed structure at 300 K. This cooling process with the intermediate relaxation simulations lasts for 312–1400 ns (depending on the conditions in Table 1); typically, the addition of C$_{60}$ results in a longer relaxation duration.

A representative P3HT:C$_{60}$ sample with DP = 150 and 50 wt% C$_{60}$ at 300 K and 1 atm is illustrated in Figure 1b. The box lengths are approximately 23.2, 24.9, 27.0, and 31.1 nm for the C$_{60}$ mass fractions of 0%, 30%, 50%, and 70%, respectively, which are of a similar order to the thickness of active layers in real OSCs.[58] The densities at 300 K and 1 atm are summarized in Table 1. It is clear that the inclusion of more C$_{60}$ molecules results in larger density values. For each C$_{60}$ mass fraction, the densities for DP = 50, 100, and 150 are very similar to each other, although the systems with larger DPs demonstrate slightly lower density values. This presumably occurs because systems with longer chains generally require a longer relaxation time for sufficient exploration of the configuration space than those with shorter chains, resulting in slightly less packed structures. The density of the pure P3HT sample (DP = 150) is approximately 0.9978 ± 0.0005 g/cm$^3$ (Table 1), showing reasonable agreement with the experimentally determined density of the amorphous phase in regioregular P3HT of similar molecular weight (1.094 g/cm$^3$).[59]

In addition to the aforementioned samples, we also construct composite systems at 300 K (i.e., with the film densification process performed at 300 K throughout the simulations). Hereafter, we refer to the systems undergoing relaxation at 600 K as "annealed" samples and to those prepared at 300 K as "unannealed" samples. The densities of the unannealed samples at 300 K and 1 atm are summarized in Table 2. Although the density differences between the annealed and unannealed samples are very small, the unannealed samples with larger DP and C$_{60}$ mass fraction values seem to have slightly lower density values, indicating that sample preparation at 300 K results in a slightly less packed structure. We use unannealed samples of 0 and 50 wt% C$_{60}$ only. We note that, for the unannealed samples, a duration of > 5 μs is recorded for the convergences of the total potential energy, radii of gyration, and end-to-end distances of the molecular chains, which is significantly longer than for the annealed samples. Unless otherwise specified, the following discussion relates to the results obtained from the annealed samples.

**Table 2. Comparison of Densities and Tensile Moduli of the Annealed and Unannealed Samples**

| DP[a] | $C_{60}$ mass fraction (wt%) | Annealing | Density (g/cm$^3$)[b] | Tensile modulus (GPa)[c] |
|---|---|---|---|---|
| 50 | 0 | Y | 0.9987 ± 0.0005 | 1.08 ± 0.04 |
|  |  | N | 0.9983 ± 0.0005 | 0.98 ± 0.03 |
|  | 50 | Y | 1.2586 ± 0.0004 | 1.96 ± 0.02 |
|  |  | N | 1.2537 ± 0.0004 | 1.92 ± 0.03 |
| 100 | 0 | Y | 0.9982 ± 0.0005 | 1.03 ± 0.03 |
|  |  | N | 0.9972 ± 0.0005 | 1.05 ± 0.08 |
|  | 50 | Y | 1.2576 ± 0.0004 | 1.97 ± 0.07 |
|  |  | N | 1.2518 ± 0.0004 | 1.82 ± 0.01 |
| 150 | 0 | Y | 0.9978 ± 0.0005 | 1.06 ± 0.05 |
|  |  | N | 0.9962 ± 0.0005 | 0.97 ± 0.08 |
|  | 50 | Y | 1.2571 ± 0.0004 | 1.98 ± 0.03 |
|  |  | N | 1.2513 ± 0.0004 | 1.83 ± 0.05 |

[a] DP refers to degree of polymerization.
[b] Densities are calculated at 300 K and 1 atm, where each uncertainty represents a standard deviation of the fluctuating density during isothermal–isobaric (NPT) MD simulations.
[c] Each uncertainty represents a standard deviation obtained from three independent simulations.

We analyze the physical conformation of the molecular chains of the annealed samples at 300 K and 1 atm by characterizing local packing structures with $d_{[001]}$ spacing in the backbone direction and $d_{[010]}$ spacing in the π-stacking direction, as illustrated in Figure 2a. Here, $d_{[010]}$ spacing is calculated by determining the proximity of the thiophene ring in each monomer (P$_1$) to others in the morphology, following the method proposed by Jones et al.[49] The $d_{[001]}$ spacings for DP = 150 are almost identical regardless of the C$_{60}$ mass fractions, as indicated in Figure 2b (and Figure S1 in Supporting Information for DP = 50 and 100). Each $d_{[001]}$ spacing is unimodal and can be fit by a Gaussian with mean $\bar{d}_{[001]}$ and standard deviation $\sigma_{[001]}$: ($\bar{d}_{[001]}$, $\sigma_{[001]}$) = (7.57 Å, 0.13 Å), (7.56 Å, 0.14 Å), (7.56 Å, 0.14 Å), and (7.54 Å, 0.14 Å) for 0, 30, 50, and 70 wt% C$_{60}$, respectively. Meanwhile, the $d_{[010]}$ spacings for DP = 150 vary for different C$_{60}$ mass fractions, as shown in Figure 2c (and Figure S1 in Supporting Information for DP = 50 and 100). The $d_{[010]}$ spacings exhibit bimodal distributions, with the first peaks characterizing crystal packing structures and the second peaks corresponding to less ordered regions of the morphologies. As indicated in Figure 2c, the presence of C$_{60}$ molecules hinders crystal packing of P3HT in the π-stacking direction. The $d_{[010]}$ spacings can be fit by two Gaussians, one having mean $\bar{d}_{[010]}$ and standard deviation $\sigma_{[010]}$ and the other having $\bar{d}_a$ and $\sigma_a$: ($\bar{d}_{[010]}$, $\sigma_{[010]}$, $\bar{d}_a$, $\sigma_a$) = (4.49 Å, 0.58 Å, 8.15 Å, 1.92 Å), (4.55 Å, 0.59 Å, 8.81 Å, 2.48 Å), (4.63 Å, 0.72 Å, 9.90 Å, 2.76 Å), and (4.87 Å, 0.77 Å, 11.58 Å, 3.00 Å) for 0, 30, 50, and 70 wt% C$_{60}$, respectively. Grazing incidence wide-angle X-ray scattering data pertinent to pure P3HT samples of similar molecular weight has been experimentally reported at $\bar{d}_{[001]}$ = 7.7 Å and $\bar{d}_{[010]}$ = 3.8 Å.[60] Our calculated value of $\bar{d}_{[001]}$ = 7.57 Å for the pure P3HT sample shows good agreement with the experimental value, whereas the calculated value of $\bar{d}_{[010]}$ = 4.49 Å is slightly larger than the experimental value. This slight difference in $\bar{d}_{[010]}$ can presumably be attributed to the approximation of the thiophene ring as a spherical site (P$_1$), resulting in overestimation of the steric bulk of the thiophene ring in the π-stacking direction.[48]

We also compare the $d_{[001]}$ and $d_{[010]}$ spacings of the annealed samples with those of the unannealed samples, as shown in Figure S2 in Supporting Information. The annealing process in the present study seems hardly to affect local packing structures, although the first peak of $d_{[010]}$ spacing for the unannealed sample of pure P3HT is slightly higher. We also note that no apparent phase separation or fullerene crystallization[61,62] takes place over the course of the sample preparation process; their time scales are far beyond those accessible to MD simulations. Nevertheless, the annealing process serves to enhance molecular chain entanglements in the samples, which in turn leads to an increase in tensile strength, as discussed in Section 3.2.

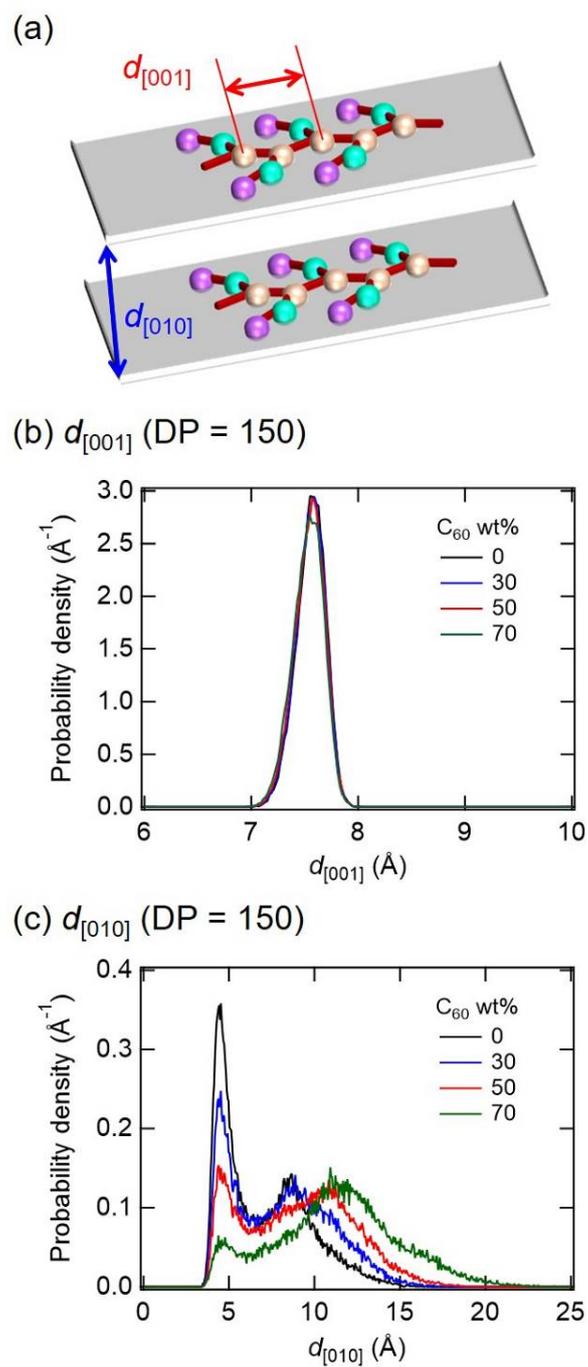

**Figure 2.** Characterization of local packing structures of the annealed samples at 300 K and 1 atm. (a) A schematic of $d_{[001]}$ spacing in the backbone direction and $d_{[010]}$ spacing in the $\pi$-stacking direction. (b) $d_{[001]}$ spacing for DP = 150 with varying $C_{60}$ mass fractions. (c) $d_{[010]}$ spacing for DP = 150 with varying $C_{60}$ mass fractions.

## 2.3. Uniaxial Tensile Simulations.

Uniaxial tensile simulations are performed with an engineering strain rate of $10^{-4}$ $ps^{-1}$ to obtain stress–strain responses for the P3HT:$C_{60}$ samples, as shown in Figure 1c. This strain rate is chosen based on our preliminary observation that strain rates of $10^{-5}$ $ps^{-1}$ and $10^{-4}$ $ps^{-1}$ result in similar stress–strain responses and almost identical tensile moduli, whereas the strain rate of $10^{-3}$ $ps^{-1}$ leads to a slightly larger tensile modulus in accordance with a previous study.[36] The strain rate of $10^{-4}$ $ps^{-1}$ is hence selected as a satisfactory compromise. Admittedly, this strain rate is orders of magnitude larger than typical strain rates used in experiments,[63,64] which are unfortunately inaccessible to MD simulations. Nevertheless, changes occur to some degree in the chain conformation during tensile deformation; thus, it is useful and fair to compare the stress–strain responses of different samples as long as the same strain rate is applied. In this study, all the uniaxial tensile simulations are performed at 300 K, with the pressure perpendicular to the tensile direction maintained at 1 atm. The tensile simulations are performed in all three directions to obtain statistical uncertainties. We note that craze formation has previously been experimentally observed for the strain of around 2% for a P3HT:PCBM sample,[16] while all the tensile simulations exhibit craze formation at much larger strains (> 200%) in the present study. This difference originates from the presence of defects at the surface that initiate craze formation in experimental conditions, whereas CGMD systems do not have surfaces due to the imposed periodic boundary conditions. Additionally, it has been reported that fracture occurs by chain pullout in low-molecular-weight P3HT samples (15 kDa), while fracture in high-molecular-weight samples (> 25 kDa) is mainly due to chain scission.[38] Although the present CG simulations (< 25 kDa) do not treat chain scission, we expect that they can reasonably represent the effect of chain pullout on the mechanical properties of bulk P3HT:$C_{60}$ samples over the course of tensile deformation.

## 3. RESULTS AND DISCUSSION

### 3.1. Stress–Strain Responses.

Stress–strain responses for the pure P3HT samples with different DPs are presented in the top panel of Figure 3a. The stress–strain curves exhibit linear elastic response for small strain ($\lesssim$ 0.02), followed by yielding and subsequent strain hardening until the stress reaches a maximum value, which is referred to as the tensile strength. The tensile moduli calculated from the linear elastic regimes (axial strain < 0.02) are 1.08 ± 0.04, 1.03 ± 0.03, and 1.06 ± 0.05 GPa for DP = 50, 100, and 150, respectively, as summarized in Table 1. These values are very similar to each other with overlapping error bars, and the value of 1.06 ± 0.05 GPa for DP = 150 is in excellent agreement with an experimentally determined value of 1.09 ± 0.15 GPa for as-cast thin films of similar molecular weight.[65] Insensitivity of the tensile modulus to DP was also reported in a previous study,[30] where elastic behavior of P3HT was observed for DP > 50. Meanwhile, our findings indicate that tensile strength (defined as maximum stress during tensile deformation) significantly increases with DP. We will return to this point in relation to molecular chain entanglements in Section 3.2.

Stress–strain responses for the P3HT:$C_{60}$ samples of DP = 100 and 150 are provided in the top panels of Figure 3b and c, with those for DP = 50 provided in the top panel of Figure S3 in Supporting Information. The results indicate that the addition of $C_{60}$ leads to an increase in the tensile modulus, indicating anti-plasticizing effects, as summarized in Table 1. Again, DP seems hardly to affect the tensile moduli of the samples. Although direct comparison is not possible due to the lack of experimental data pertinent to P3HT:$C_{60}$ samples, a previous study reported the tensile modulus of an as-cast thin film of a 1:1 P3HT:PCBM mass ratio to be 1.97 ± 0.07 GPa,[66] which is very close to our calculated value (1.98 ± 0.03 GPa for DP = 150 and 50 wt% $C_{60}$). Additionally, our findings show that the tensile strength tends to decrease with increasing $C_{60}$ mass fraction for DP = 100 and 150, whereas there is no clear relationship between the tensile strength and the $C_{60}$ mass fraction for DP = 50 (Figure S3).

The dependence of stress–strain responses on the annealing process is shown for the pure P3HT samples in Figure 4 and for the P3HT:$C_{60}$ samples with 50 wt% $C_{60}$ in Figure S4 of Supporting Information. Although the differences between the tensile moduli of the annealed and unannealed samples are small, with overlapping error bars in some cases, the annealed samples (especially the P3HT:$C_{60}$ composites) seem to have slightly larger tensile modulus values, as summarized in Table 2. This slight increase in the tensile modulus has a positive correlation with the density, where the annealed samples have slightly larger density values. These results indicate that the annealing process leads to slightly more packed structures, which in turn slightly increases the tensile modulus, although the local packing structures of the annealed samples characterized by $d_{[001]}$ and $d_{[010]}$ spacings remain very close to those of the unannealed samples, as evidenced in Figure S2 in Supporting Information. Furthermore, we observe that the stress–strain responses of the annealed and unannealed samples with larger DP values differ significantly beyond the linear elastic regimes. Specifically, the tensile strengths of the annealed samples are larger than those of the unannealed samples, and the difference between the tensile strengths of the annealed and unannealed samples increases with DP.

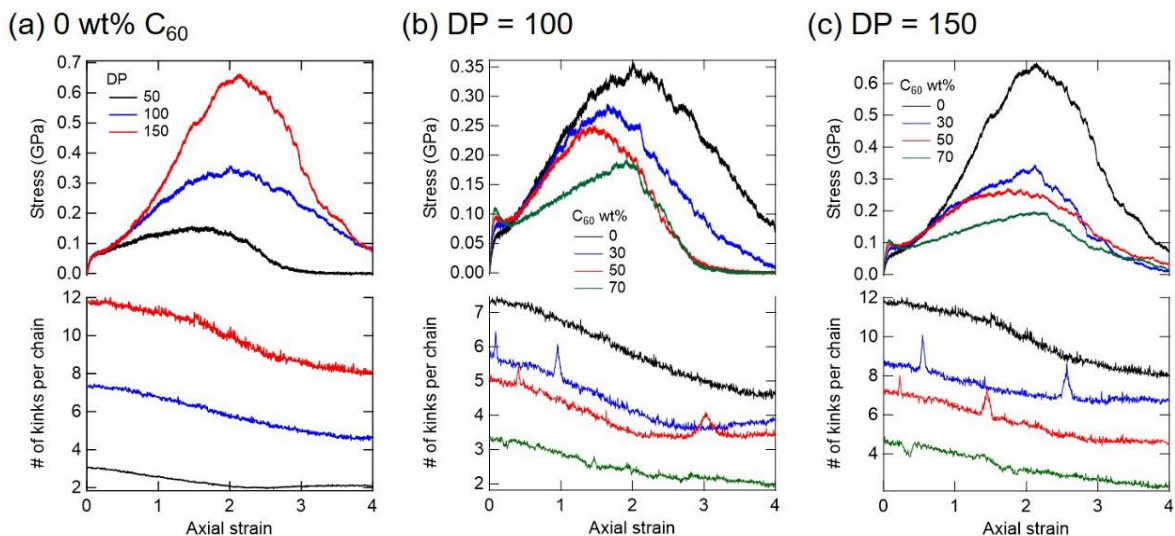

**Figure 3.** Stress–strain responses (top) and average number of kinks per chain (bottom) for the annealed samples. (a) DP dependence for 0 wt% $C_{60}$. (b) $C_{60}$ mass fraction dependence for DP = 100. (c) $C_{60}$ mass fraction dependence for DP = 150.

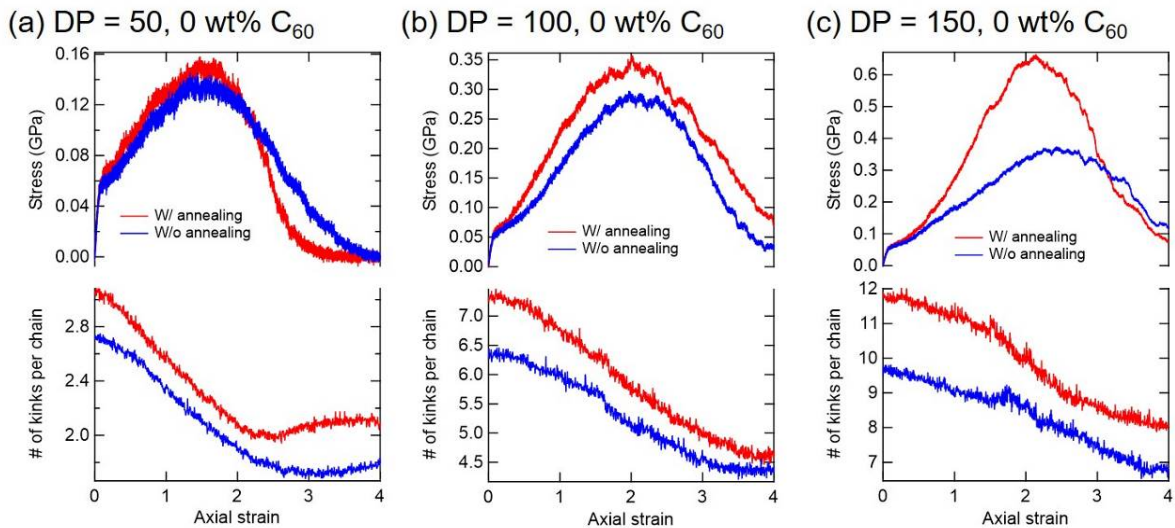

**Figure 4.** Stress–strain responses (top) and average number of kinks per chain (bottom). Annealing dependence for the pure P3HT samples is shown for DP = 50 (a), 100 (b), and 150 (c), respectively.

## 3.2. Molecular Chain Entanglements.

From the results presented in Section 3.1, we find that the inclusion of $C_{60}$ results in a larger tensile modulus value and tends to decrease tensile strength. Additionally, the annealing process leads to an increase in tensile strength, especially for larger DPs, whereas it seems to have only a minor effect on the tensile modulus. Essentially, the tensile strength of a sample is located at a large strain regime and hence is closely related to molecular chain configuration, whereas the tensile modulus characterizes a linear elastic regime at small strain values, primarily originating from molecular interactions at atomistic levels rather than molecular chain configuration at chain-length levels (see Section 3.3).

To rationalize these tendencies, we quantify molecular chain entanglements based on primitive path (PP) analysis,[41–44] a schematic for which is depicted in Figure 5. In the primitive analysis of a polymeric system, the ends of each molecular chain are fixed in space, with each molecular chain represented by multiple nodes. Then, the length of the multiple disconnected path is monotonically reduced through geometrical transformation; this results in a smaller number of nodes while adhering to chain uncrossability. By iterating this procedure, the system eventually converges at a final state (i.e., the shortest path). The PP is defined as the shortest path that connects the two ends of the molecular chain, and kinks are defined as the crossing points of the PPs (as illustrated in Figure 5). Therefore, the PP analysis allows for the extraction of inherent molecular chain entanglements from polymeric systems. In the present study, we evaluate the average number of kinks per chain as a function of the axial strain using the Z1 code developed by Kröger and coworkers.[41–44]

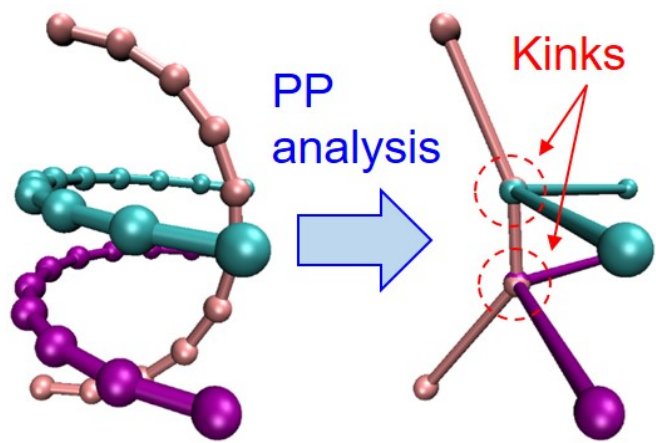

**Figure 5.** Schematic of the primitive path (PP) analysis—the length of the multiple disconnected path is monotonically reduced, with the chain ends fixed in space, subject to chain uncrossability, until the shortest path is obtained.

Figure 6 shows the average number of kinks per chain, $\langle Z \rangle$, for the unstrained systems (axial strain = 0). $\langle Z \rangle$ increases with DP regardless of the $C_{60}$ mass fractions, as shown Figure 6a, and the bottom panel of Figure 3a clearly illustrates decreases in $\langle Z \rangle$ with increasing strain, thus successfully characterizing the disentanglements of the molecular chains. Furthermore, Figure 3a indicates that samples with larger initial values of $\langle Z \rangle$ possess higher tensile strength. Additionally, $\langle Z \rangle$ decreases with an increase in the $C_{60}$ mass fraction for each DP, as shown in Figure 6a. P3HT:$C_{60}$ samples of DP = 100 or 150 with larger $C_{60}$ mass fraction demonstrate smaller $\langle Z \rangle$ values and tend to exhibit lower tensile strength, as indicated in the bottom panels of Figure 3b and c. These results indicate that molecular chain entanglement is closely related to tensile strength. This is further supported by Figure 4, where the difference between the values of $\langle Z \rangle$ for the annealed and unannealed samples correlates well with the difference in tensile strengths between them. We note that the difference of $\langle Z \rangle$ between the annealed and unannealed samples tends to increase with DP, as summarized for 0 wt% $C_{60}$ in Figure 6b and for 50 wt% $C_{60}$ in Figure S5 in Supporting Information. This is because sufficient exploration of the configuration space during the densification process could not be achieved in the unannealed samples with large DP values (which was not the case in the annealed samples), resulting in a system with less entangled molecular chains. We also find that, for the P3HT:$C_{60}$ samples of DP = 50, $\langle Z \rangle$ does not seem to have an apparent correlation with the tensile strength, as shown in Figure S3 in Supporting Information. This is presumably due to the small $\langle Z \rangle$ value for DP = 50, especially in the presence of $C_{60}$ molecules, which could indicate that the molecular chains do not exhibit truly entangled behavior[30] and that $\langle Z \rangle$ is not a determinant of tensile strength.

Figure 4 and Table 2 indicate that the tensile modulus is not sensitive to the molecular chain entanglements. Figure 3 and Table 1 also exhibit little dependence of the tensile modulus on DP and hence on the chain entanglements. Meanwhile, in a previous study,[32] all-atom MD simulations of donor−acceptor polymers (PDTSTPD, PTB7, and TQ1) showed a drastic decrease in the tensile modulus in a self-aggregated morphology of lower density relative to a melt-quenched morphology of higher density. From these observations, we expect that molecular interactions are a more direct determinant of the tensile modulus than the chain entanglements. In this regard, we further investigate the molecular origin of the tensile modulus in Section 3.3.

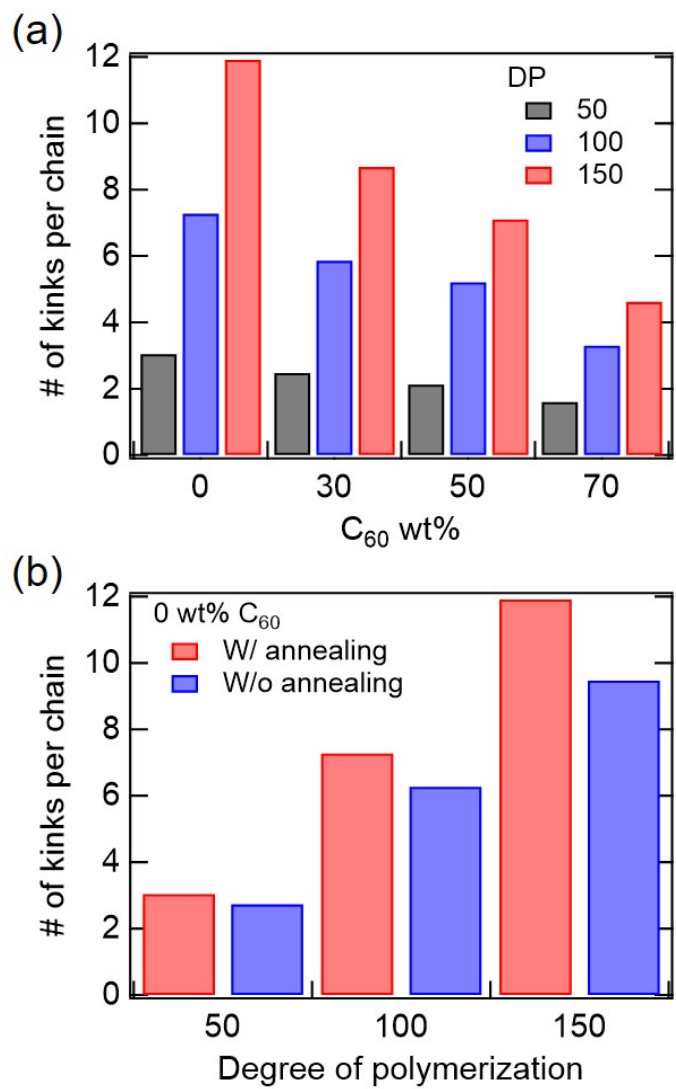

**Figure 6.** Average number of kinks per chain for the unstrained systems. (a) Summary for the annealed samples of DP = 50, 100, and 150 with 0, 30, 50, and 70 wt% $C_{60}$, respectively. (b) Comparison of the annealed and unannealed samples for 0 wt% $C_{60}$.

3.3. Molecular Interactions Contributing to Tensile Modulus.

To elucidate the origin of the tensile modulus, we examine the molecular interactions contributing to the total stress. The stress tensor comprises a kinetic energy contribution and a virial contribution resulting from intra- and inter-molecular interactions,[28] which are represented by non-bonded interactions and by bonded interactions involving bond-stretching, angle-bending, and dihedral and improper potentials, as described in Section 2.1. To decompose the stress tensor, per-particle stress tensor values are first calculated for each contribution;[67] then, these values are summed over all the CG particles to obtain the stress–strain responses decomposed into each contribution, as illustrated in Figure S6 in Supporting Information. Their contributions to the tensile modulus (calculated from the linear elastic regimes) are summarized for DP = 150 in Figure 7a and for DP = 50 and 100 in Figure S7 in Supporting Information. As a general tendency, the tensile modulus is shown to be primarily determined by non-bonded interactions and by bonded interactions associated with bond-stretching and angle-bending, whereas bonded interactions associated with dihedral and improper potentials seem to have a negligibly small impact on the tensile modulus. Interestingly, for the pure P3HT sample, the results indicate that bond-stretching potential makes the largest contribution to the tensile modulus, with a comparable contribution from the non-bonded interactions, indicating elastic energy being mainly stored by the non-bonded interactions and the bonded interactions associated with bond-stretching. An enhanced contribution from non-bonded interactions is seen with the addition of $C_{60}$ molecules, leading to these interactions becoming a primary factor in the tensile modulus for $C_{60}$ mass fractions ≥ 50 wt%, while the contribution from bond-stretching exhibits less dependence on the $C_{60}$ mass fraction. The non-bonded interactions contributing to the tensile modulus are further decomposed into P3HT–P3HT, P3HT–$C_{60}$, and $C_{60}$–$C_{60}$ interactions for DP = 150 (Figure 7b) and for DP = 50 and 100 (Figure S7 in Supporting Information). The non-bonded interactions associated with $C_{60}$ (P3HT–$C_{60}$ and $C_{60}$–$C_{60}$) increase with the $C_{60}$ mass fraction, whereas the P3HT–P3HT interactions decrease as the $C_{60}$ mass fraction increases. These results indicate that the increases in the tensile modulus when $C_{60}$ molecules are added mainly originate from an enhanced contribution from the non-bonded interactions associated with $C_{60}$.

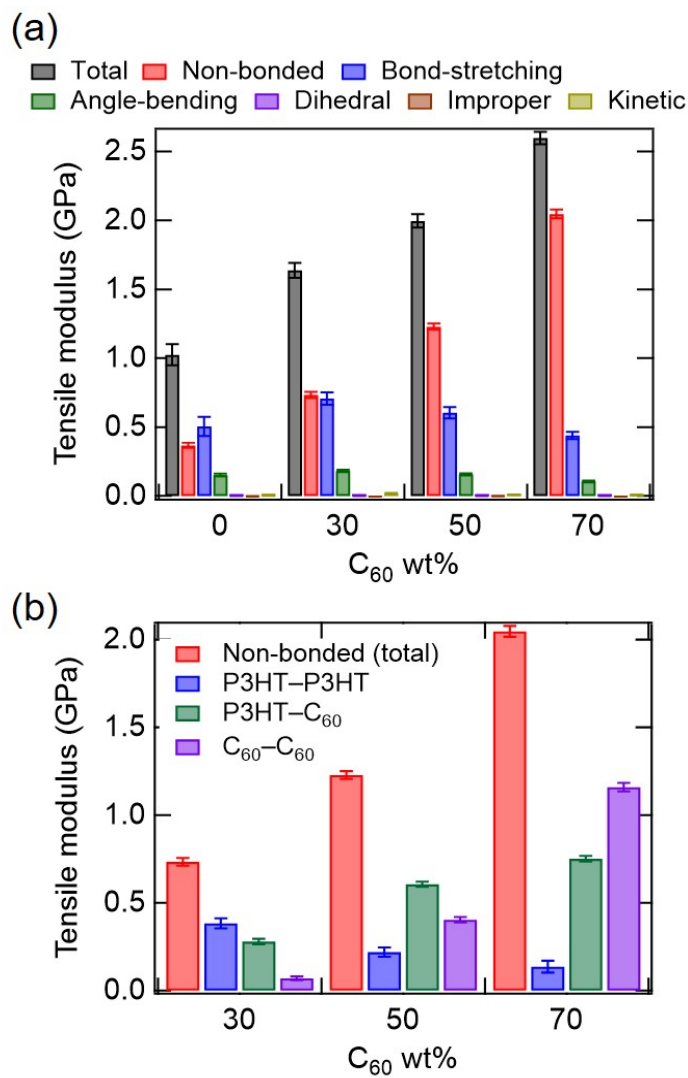

**Figure 7.** Decomposition of the tensile modulus for DP = 150. (a) Contributions to the tensile modulus from kinetic energy, non-bonded interactions, and bonded interactions (bond-stretching, angle-bending, and dihedral and improper potentials) for different $C_{60}$ mass fractions. Each uncertainty represents a standard deviation associated with least squares fitting (see Figure S6 in Supporting Information). (b) The non-bonded interactions contributing to the tensile modulus are further decomposed into P3HT–P3HT, P3HT–$C_{60}$, and $C_{60}$–$C_{60}$ interactions.

### 3.4. Persistent Homology Analysis.

During tensile deformation, strained samples with higher $C_{60}$ mass fractions exhibit more pronounced yield behaviors before strain hardening, as shown in the top panels of Figure 10. Here, in terms of solid mechanics, an upper yield point refers to the point at which the stress arrives at a local maximum value after an elasticity limit, followed by a decrease in the stress that characterizes strain softening. After the strain-softening regime, the stress arrives at a local minimum value, which is referred to as a lower yield point. After the system arrives at the lower yield point, it undergoes strain hardening. As shown in the top panels of Figure 10, the pure P3HT sample does not exhibit apparent upper and lower yield points, whereas these points are more pronounced for the P3HT:$C_{60}$ samples with higher $C_{60}$ mass fractions. In the present study, we relate the emergence of the upper and lower yield points to the microvoids generated in the course of tensile deformation based on persistent homology analysis,[45,46] which can extract multiscale topological features embedded in a molecular system. A schematic of the persistent homology analysis is depicted in Figure 8. Each CG particle possesses the input radius $r_i$ ($i = 1, ..., N$), where $N$ is the number of CG particles in the system. In this study, the $r_i$ values for each particle are set as Lennard-Jones length parameters.[36,49] Then, by introducing a resolution parameter $\alpha$, the particle radius is rewritten as

$$r_i(\alpha) = \sqrt{r_i^2 - r_{\min}^2 + \alpha} \tag{1}$$

$$r_{\min} = \min\{r_1, ..., r_N\} \tag{2}$$

The particle radii are varied through $\alpha$ ($> 0$), and cavities are detected at each $\alpha$. More specifically, as illustrated in Figure 8, by gradually increasing $\alpha$, a void surrounded by certain particles is generated at $\alpha = \alpha_b$, which is referred to as the birth scale of the void. With further increases of the particle radii, the void vanishes at $\alpha = \alpha_d$, which is referred to as the death scale of the void. Thus, the death scale is a semiquantitative measure of the void size, although the definition of the void is ambiguous due to its anisotropic nature. The definition of the particle radii shown in Equation (1) ensures a Voronoi tessellation of the system which is invariant to changing $\alpha$. Subsequently, by plotting birth–death scales of the voids as a 2D histogram, known as a persistent diagram (Figure 9), it is possible to visualize multiscale topological features embedded in the molecular system. Existing sources present mathematically rigorous details of the persistent homology analysis.[45,68] The persistent homology analysis for the current study is performed using HomCloud.[69]

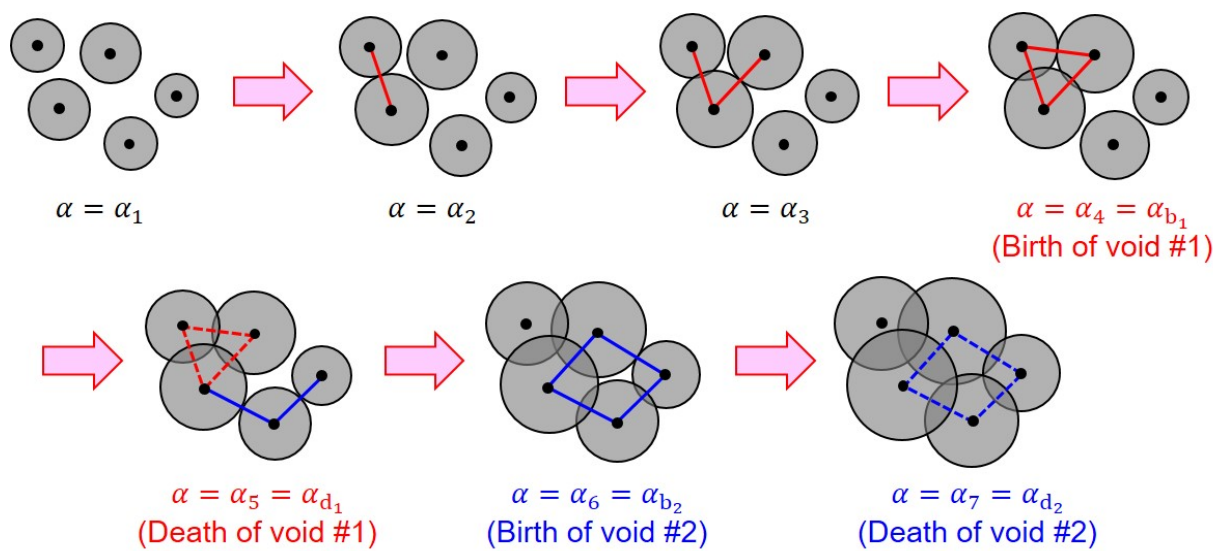

**Figure 8.** A schematic of the persistent homology analysis. Each CG particle possesses a radius (defined by Equation (1)). By gradually increasing the particle radii, a void surrounded by certain particles is generated at $\alpha = \alpha_b$ (the birth scale of the void). As particle radii increase further, the void vanishes at $\alpha = \alpha_d$ (the death scale of the void).

Figure 9a shows the persistent diagrams for the P3HT sample of DP = 150 with varied axial strains. We find that a significant number of points appear in the vicinity of the diagonal lines ($\alpha_b = \alpha_d \lesssim 15$ Å$^2$). These points represent voids with death scales that are very close to their birth scales, meaning that these voids readily vanish due to thermal fluctuation and hence represent topological noises. Therefore, these voids should be considered as physically insignificant.[46] Furthermore, the points indicating very large birth scales are also physically irrelevant to voids, as voids are surrounded by CG particles and hence on the order of CG particle sizes. In light of these factors, we assume that the region $\alpha_b < 25$ Å$^2 < \alpha_d$ corresponds to physically significant voids. Here, the lower bound of $\alpha_d$ is determined such that the larger voids appearing at the strained states are mostly detected (see Figure 9). We have also confirmed that slight modifications of these thresholds do not alter the qualitative results discussed below.

Void birth scales tend to decrease with increasing $C_{60}$ mass fraction, as shown in Figure 9. The voids surrounded by closely packed particles tend to exhibit small birth scales, whereas those surrounded by less closely packed particles demonstrate larger birth scales (Figure 8). This indicates that the addition of $C_{60}$ results in more closely packed structures. Importantly, Figure 9a shows that the number of physically significant voids ($\alpha_b < 25$ Å$^2 < \alpha_d$) increases with the axial strain, as further evidenced in the bottom panel of Figure 10a, where the number of physically significant voids generated in the course of tensile deformation is provided. For the pure P3HT samples, the number of voids increases almost linearly in the course of tensile deformation up to the axial strain of $\varepsilon \approx 0.3$, followed by a more gradual increase in the number of voids beyond this point. In contrast, for the P3HT:$C_{60}$ samples of 30 and 50 wt% $C_{60}$, the numbers of voids increase up to $\varepsilon \approx 0.22$ and 0.26, respectively, and saturation is likely to be reached beyond those strains, as shown in the bottom panels of Figure 10b and c. This saturation is shown to be closely related to the coalescence of microvoids in the course of tensile deformation, which results in the emergence of larger voids. The emergence of larger voids is also confirmed by the findings in Figure 9b and c, where larger voids with larger death scales are indicated for $\varepsilon = 1.0$ than $\varepsilon = 0.2$. The coalescence of microvoids leading to the saturation of the number of voids around $\varepsilon \approx 0.22$ and 0.26 for 50 and 70 wt% $C_{60}$, respectively, correlates well with the lower yield points shown in Figure 10b and c, indicating that the emergence of the upper and lower yield points is closely related to the coalescence of microvoids during tensile deformation. Previous studies[46,70] have reported that craze formation in glassy polymers is closely related to void coalescence. Our results indicate that yield behavior characterized by upper and lower yield points, which appears at far smaller strain than craze formation, is also closely related to the coalescence of microvoids generated in the course of tensile deformation.

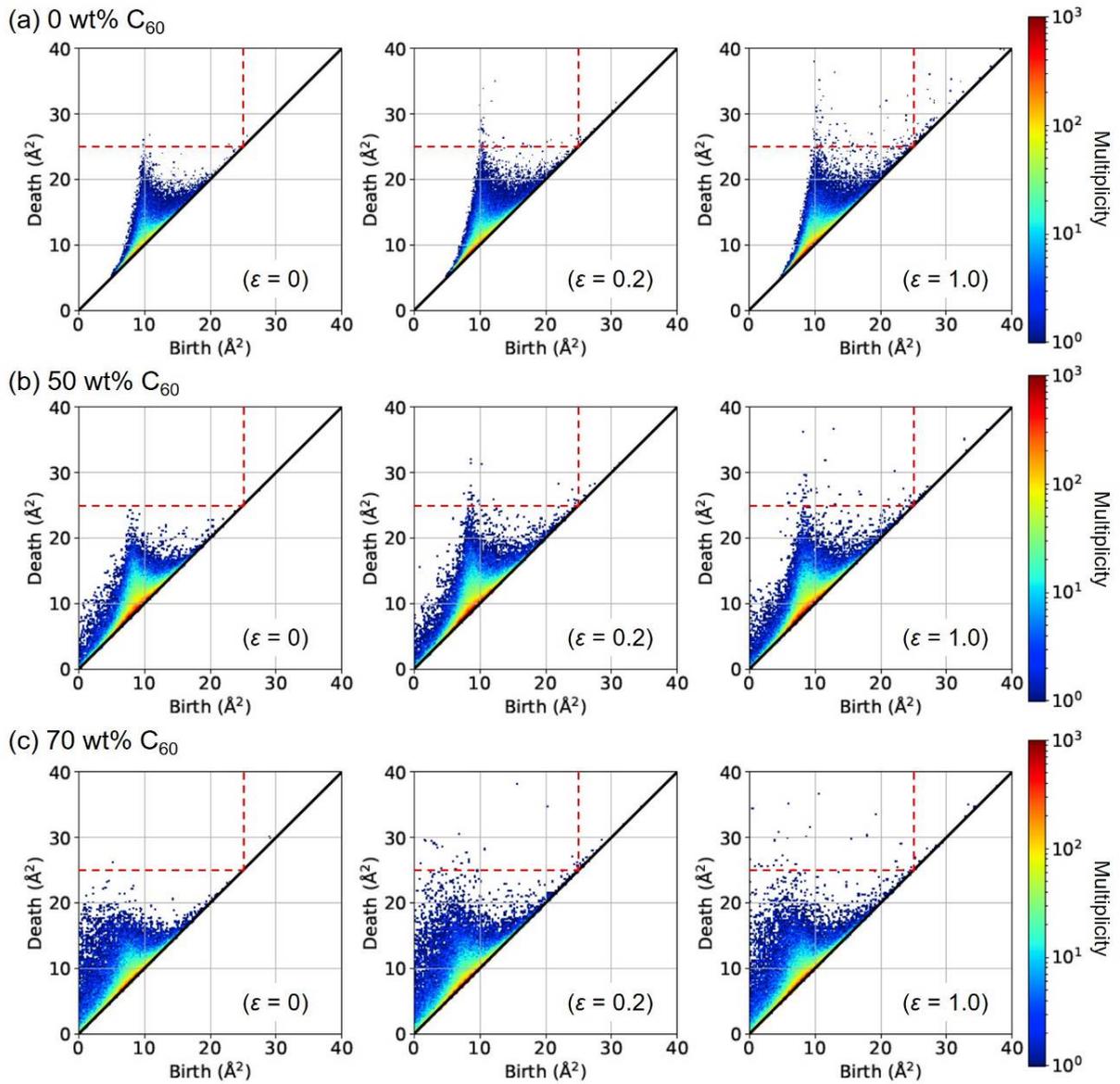

**Figure 9.** Persistent diagrams for DP = 150 with 0 wt% $C_{60}$ (a), 50 wt% $C_{60}$ (b), and 70 wt% $C_{60}$ (c) for the axial strains $\varepsilon$ = 0, 0.2, and 1.0, with the multiplicity on the logarithmic scale. The region $\alpha_b$ < 25 Å$^2$ < $\alpha_d$, represented by the dashed lines, is defined to capture physically significant voids. Numbers of physically significant voids at $\varepsilon$ = 0, 0.2, and 1.0 are as follows, respectively: (a) 5, 48, and 93 (0 wt% $C_{60}$); (b) 0, 25, and 37 (50 wt% $C_{60}$); and (c) 1, 27, and 33 (70 wt% $C_{60}$).

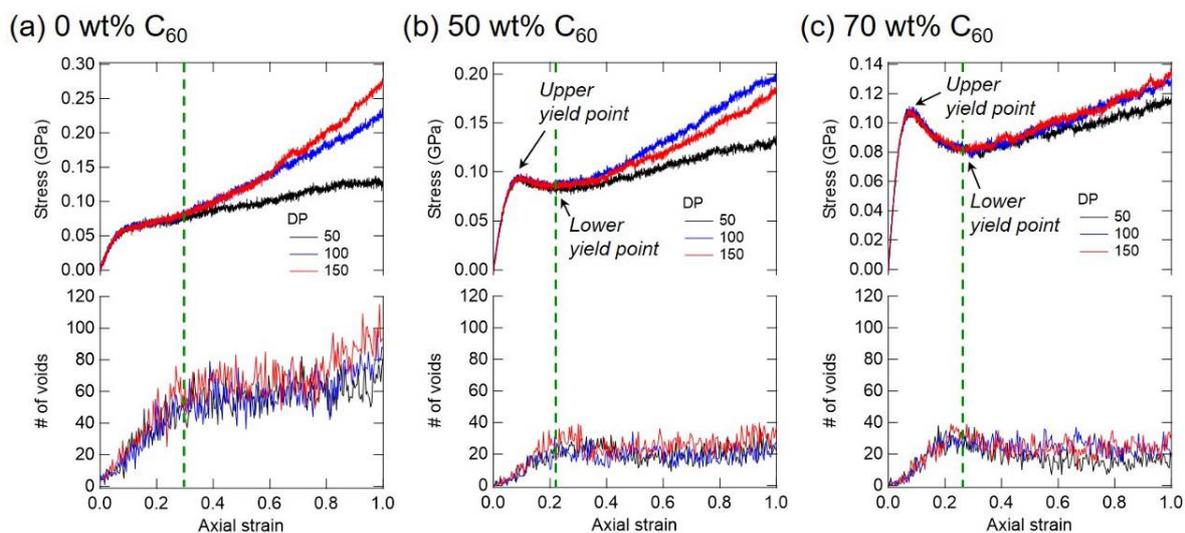

**Figure 10.** Stress–strain responses (top) and the number of physically significant voids ($\alpha_b < 25$ Å$^2$ < $\alpha_d$; see Figure 9) generated in the course of tensile deformation (bottom). The dashed lines represent strains where the numbers of voids start to exhibit different tendencies. (a) 0 wt% $C_{60}$, (b) 50 wt% $C_{60}$, and (c) 70 wt% $C_{60}$.

## 4. CONCLUSIONS

In this study, we obtain molecular insight into the mechanical properties of P3HT:$C_{60}$ composites using CGMD simulations. Decomposition of stress tensor into kinetic energy and virial contributions indicates that the tensile moduli of the pure P3HT samples are greatly dependent on non-bonded interactions and on bonded interactions associated with bond-stretching, whereas the addition of $C_{60}$ leads to an increase in the tensile modulus originating from enhanced non-bonded interactions associated with $C_{60}$. Additionally, we find that the tensile strength of the P3HT:$C_{60}$ samples correlates well with molecular chain entanglements, characterized by the average number of kinks per chain obtained from PP analysis. Finally, we find that the upper and lower yield points characterizing strain softening are more pronounced as $C_{60}$ mass fraction increases. Persistent homology analysis indicates that the emergence of the upper and lower yield points correlates well with the coalescence of microvoids in the course of tensile deformation, resulting in the generation of larger voids. The present study provides a fundamental understanding of the molecular determinants of the mechanical properties of π-conjugated polymer–fullerene composites, which can also help to interpret and predict the mechanical properties of other polymer composites involving fullerene from a microscopic perspective.

In this study, no apparent phase separation or fullerene crystallization takes place during the film preparation process. In future work, we plan to investigate how the interplay between amorphous and partially crystalline regions affects the mechanical properties of π-conjugated materials, given that actual BHJ active layers form phase-segregated, bicontinuous interpenetrating networks. It would also be of interest to simulate $P3HT:C_{60}$ samples of even higher molecular weight with polydispersity in terms of organic photovoltaic applications, although the computational burden would drastically increase for such cases. Furthermore, in actual OSC devices, surface effects play a key role in initiating craze formation, which should also be investigated in future research.

## ASSOCIATED CONTENT

**Supporting Information**

The following files are available free of charge.

Additional results related to local packing structures of $P3HT:C_{60}$ samples, the relationship between stress–strain responses and molecular chain entanglements, and the decomposition of molecular interactions into the kinetic energy and virial contributions (PDF)

AUTHOR INFORMATION

**Corresponding Author**

*E-mail yyoshimoto@fel.t.u-tokyo.ac.jp (Y.Y.).

**Notes**

The authors declare no competing financial interest.

ACKNOWLEDGMENT

This work was partly supported by Research and Education Consortium for Innovation of Advanced Integrated Science (CIAiS). The authors would like to thank Prof. Martin Kröger for providing us with an executable of the Z1 code.

# Supporting Information

# Molecular Insights into the Mechanical Properties of Polymer–Fullerene Bulk Heterojunctions for Organic Photovoltaic Applications

*Yuta Yoshimoto,\* Sou Sugiyama, Shuntaro Shimada, Toshihiro Kaneko, Shu Takagi, and Ikuya Kinefuchi*

Department of Mechanical Engineering, The University of Tokyo, 7-3-1 Hongo, Bunkyo-ku, Tokyo 113-8656, Japan

## S1. LOCAL PACKING STRUCTURES OF P3HT:$C_{60}$ SAMPLES

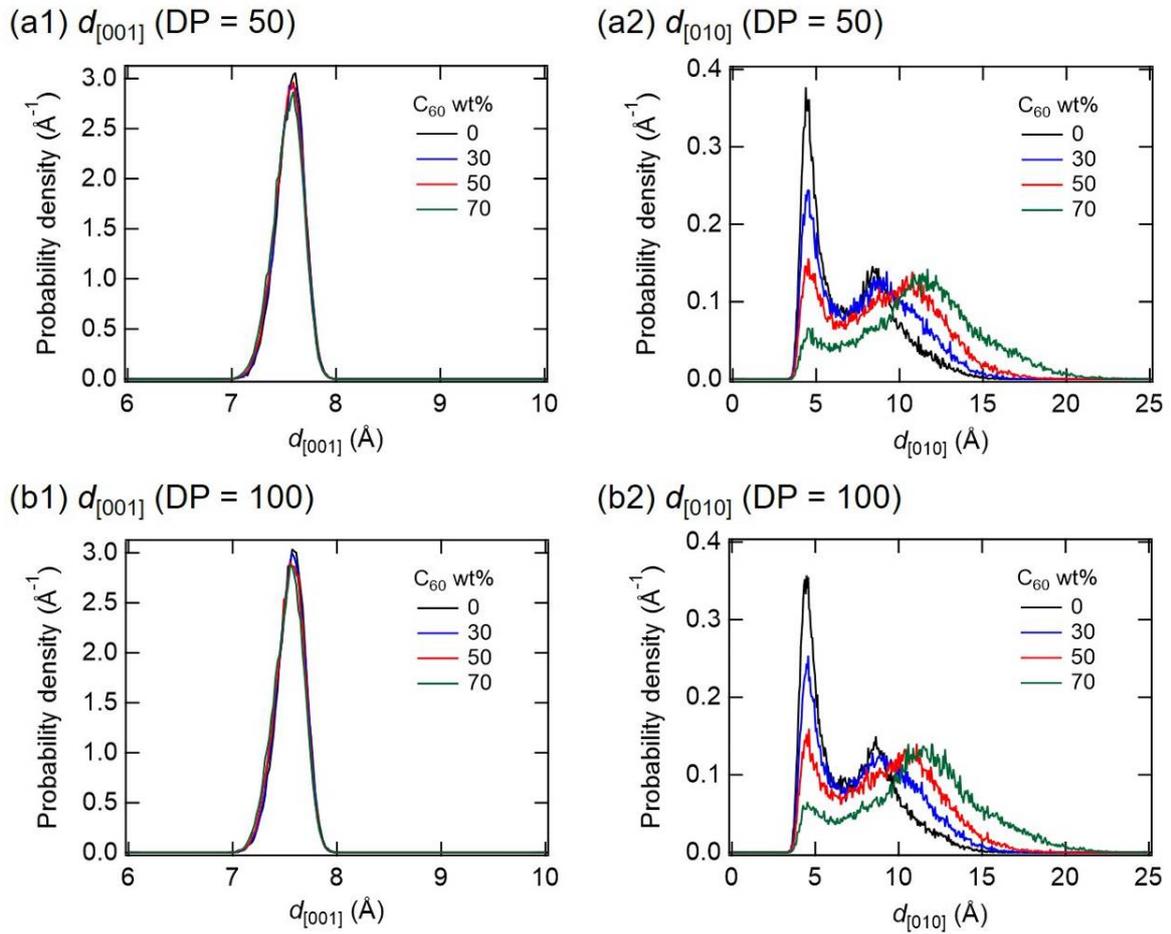

**Figure S1.** Characterization of local packing structures of the annealed samples at 300 K and 1 atm as a function of the $C_{60}$ mass fraction. (a1) $d_{[001]}$ spacing for DP = 50. (a2) $d_{[010]}$ spacing for DP = 50. (b1) $d_{[001]}$ spacing for DP = 100. (b2) $d_{[010]}$ spacing for DP = 100.

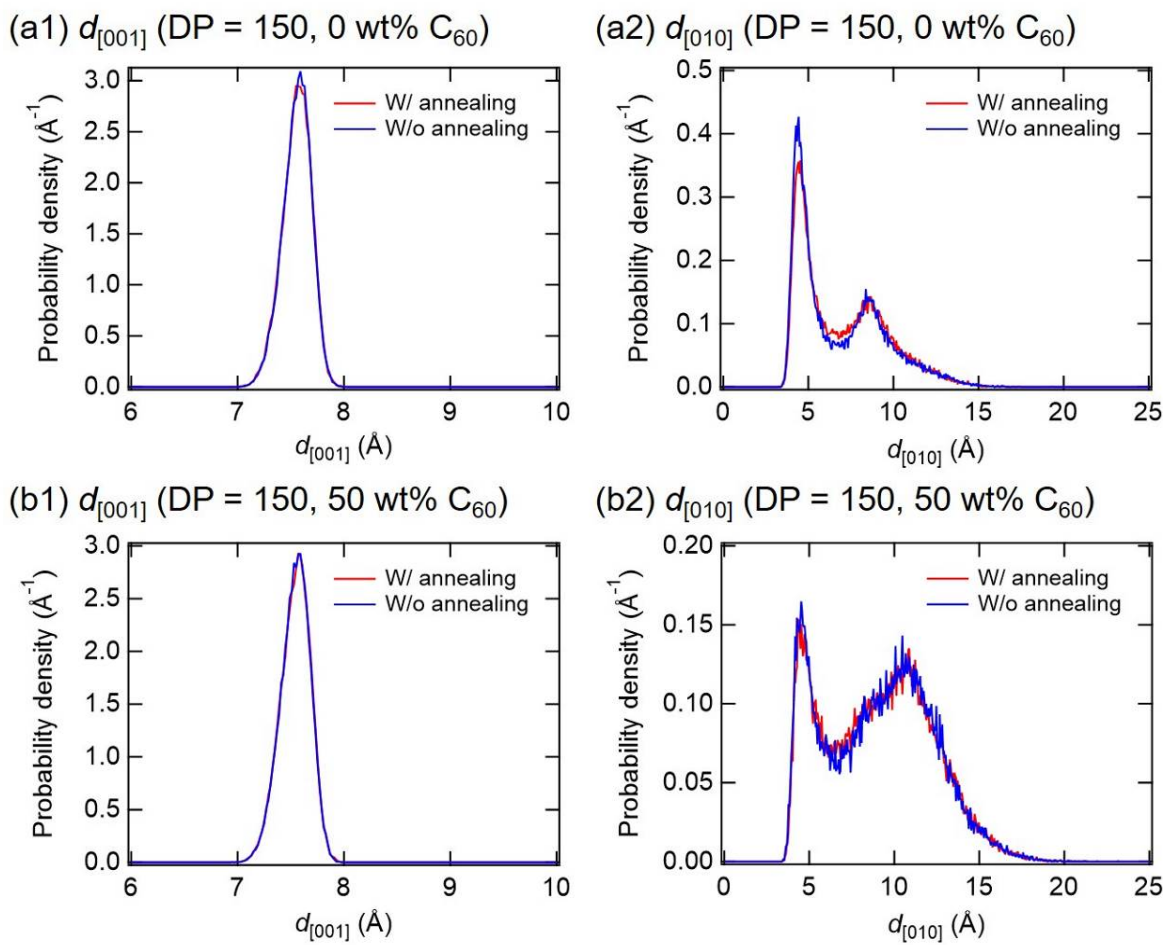

**Figure S2.** Comparison of $d_{[001]}$ and $d_{[010]}$ spacings of the annealed and unannealed samples. (a1) $d_{[001]}$ spacing for DP = 150 and 0 wt% $C_{60}$. (a2) $d_{[010]}$ spacing for DP = 150 and 0 wt% $C_{60}$. (b1) $d_{[001]}$ spacing for DP = 150 and 50 wt% $C_{60}$. (b2) $d_{[010]}$ spacing for DP = 150 and 50 wt% $C_{60}$.

## S2. STRESS–STRAIN RESPONSES AND CHAIN ENTANGLEMENTS

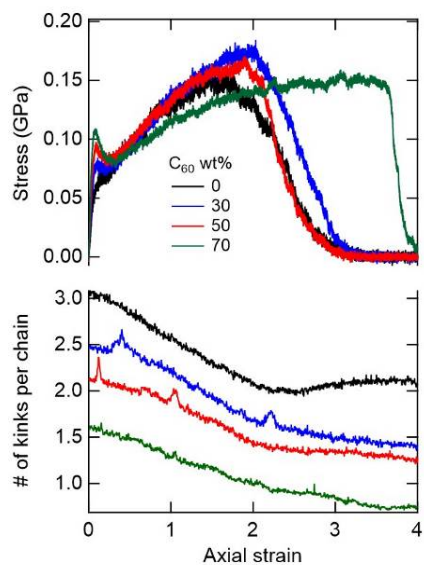

**Figure S3.** Stress–strain responses (top) and average number of kinks per chain (bottom) for the annealed samples of DP = 50 with different $C_{60}$ mass fractions.

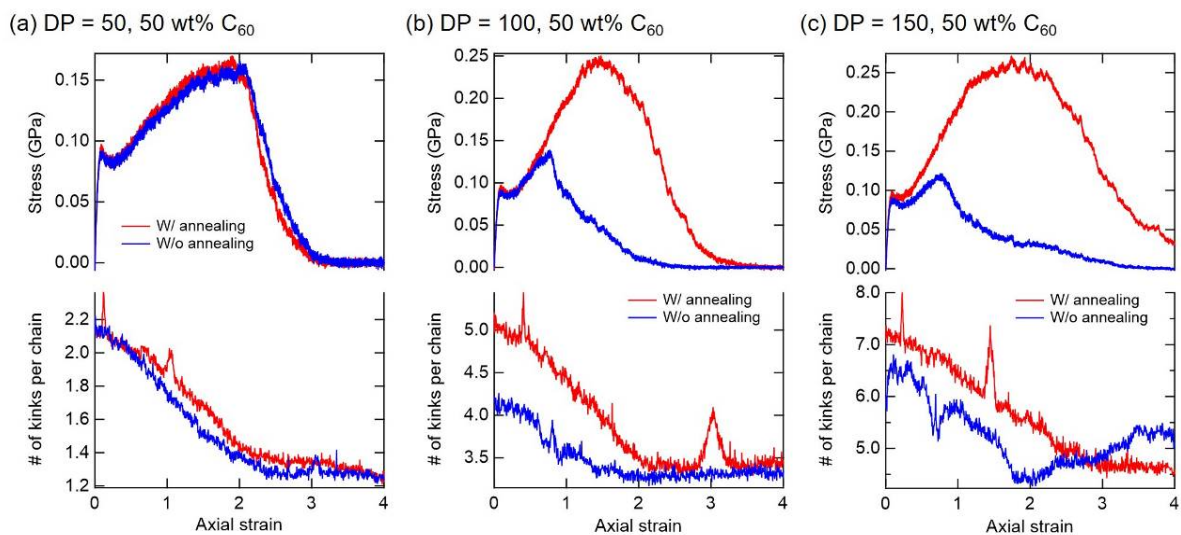

**Figure S4.** Stress–strain responses (top) and average number of kinks per chain (bottom). Annealing dependence for the P3HT:$C_{60}$ samples with 50 wt% $C_{60}$ is shown for DP = 50 (a), 100 (b), and 150 (c), respectively.

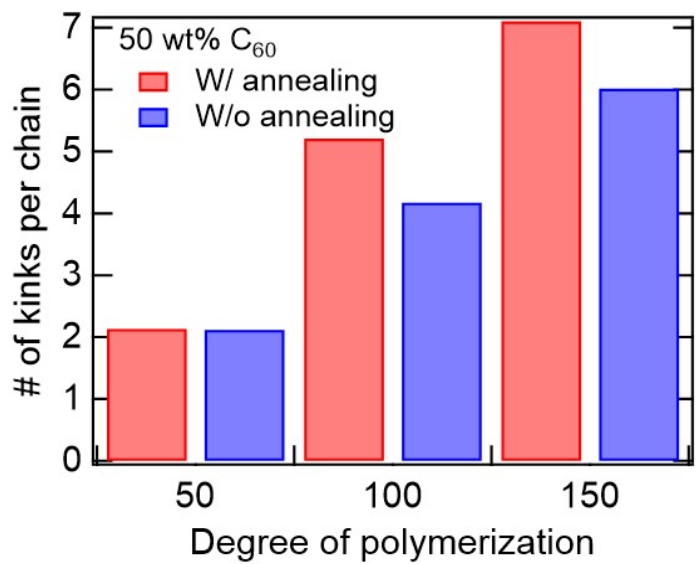

**Figure S5.** Average number of kinks per chain for the unstrained systems of the annealed and unannealed samples of 50 wt% $C_{60}$.

## S3. DECOMPOSITION OF MOLECULAR INTERACTIONS

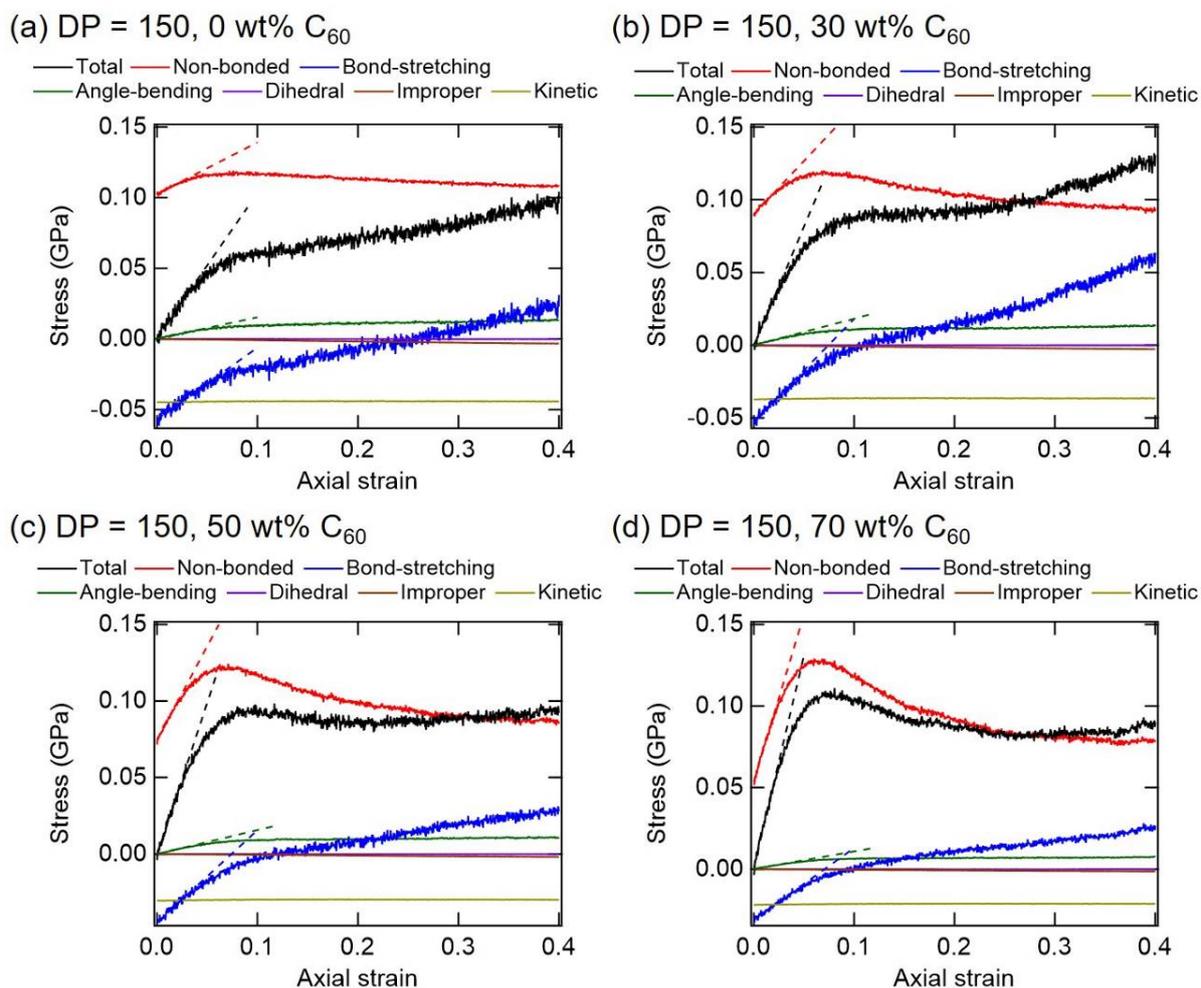

**Figure S6.** Stress tensor decomposed into the kinetic energy and the virial contributions represented by non-bonded interactions and bonded interactions (bond-stretching, angle-bending, and dihedral and improper potentials). The dashed lines represent least squares fitting to the linear elastic regimes (strain < 0.02). (a) DP = 150 and 0 wt% $C_{60}$. (b) DP =150 and 30 wt% $C_{60}$. (c) DP = 150 and 50 wt% $C_{60}$. (d) DP = 150 and 70 wt% $C_{60}$.

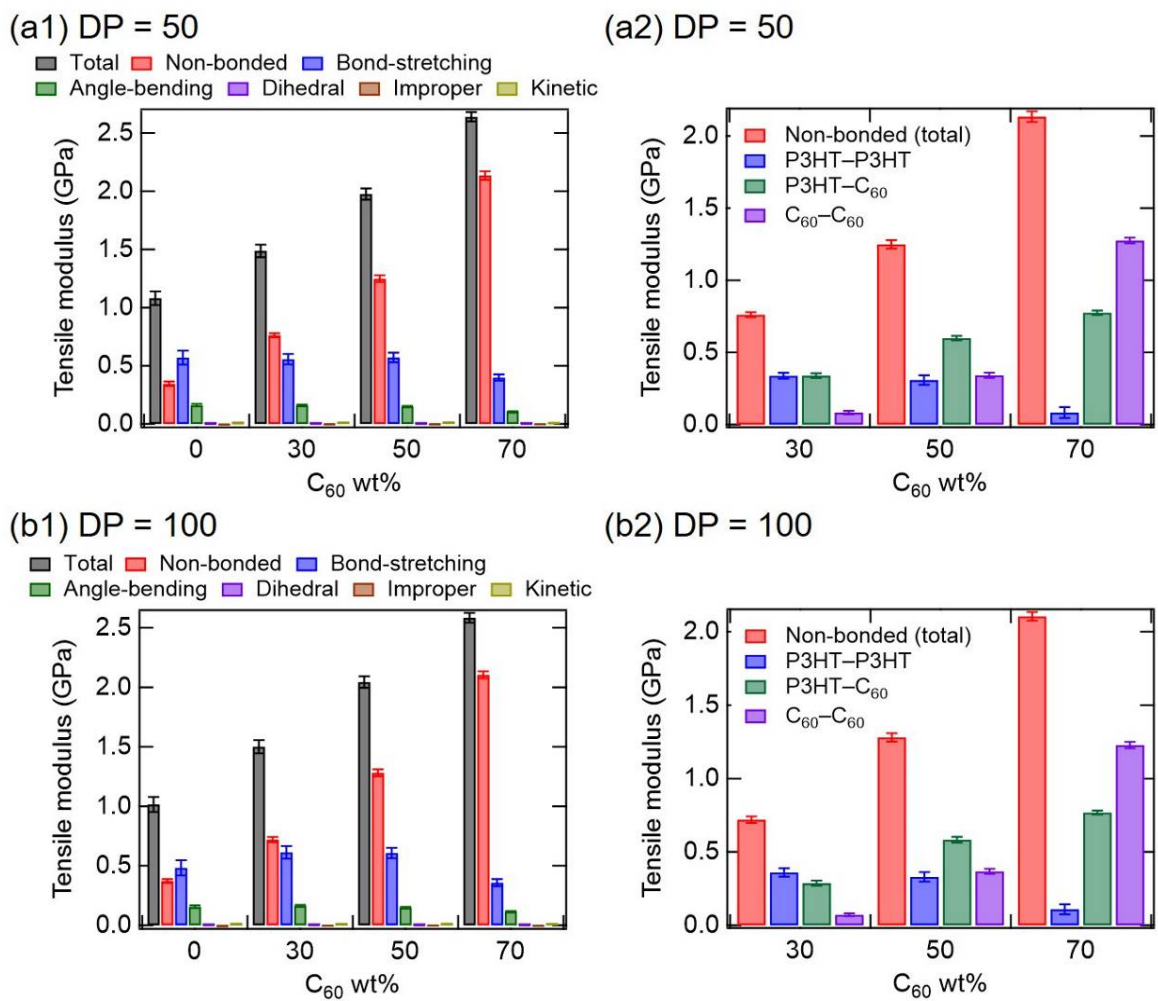

**Figure S7.** Decomposition of the tensile moduli for DP = 50 (a) and 100 (b). (a1, b1) Contributions to the tensile modulus from kinetic energy, non-bonded interactions, and bonded interactions (bond-stretching, angle-bending, and dihedral and improper potentials) for different $C_{60}$ mass fractions. Each uncertainty represents a standard deviation associated with least squares fitting. (a2, b2) The non-bonded interactions contributing to the tensile moduli are further decomposed into P3HT–P3HT, P3HT–$C_{60}$, and $C_{60}$–$C_{60}$ interactions.